\documentclass[sigconf, screen]{acmart}
\renewcommand\footnotetextcopyrightpermission[1]{} 
\AtBeginDocument{%
  }

\setcopyright{acmlicensed}
\copyrightyear{2018}
\acmYear{2018}
\acmDOI{XXXXXXX.XXXXXXX}
\acmConference[Preprint]{Make sure to enter the correct
  conference title from your rights confirmation email}{Under review}{2026}

\acmSubmissionID{8082}
\usepackage{cleveref,subcaption}

\begin{document}

\title{Reducing Linguistic Hallucination in LM-Based Speech Enhancement via Noise-Invariant Acoustic-Semantic Distillation}

\author{Zheng Wang}
\affiliation{%
  \institution{Nanjing University}
  \city{Nanjing}
  \country{China}
}
\email{zheng.wang@smail.nju.edu.cn}

\author{Xiaobin Rong}
\affiliation{%
  \institution{Nanjing University}
  \city{Nanjing}
  \country{China}
}
\email{xiaobin.rong@smail.nju.edu.cn}

\author{Hang Su}
\affiliation{%
  \institution{MiLM Plus, Xiaomi Inc.}
  \city{Beijing}
  \country{China}}
\email{suhang10@xiaomi.com}

\author{Tianyi Tan}
\affiliation{%
  \institution{Nanjing University}
  \city{Nanjing}
  \country{China}
}
\email{tianyi.tan@smail.nju.edu.cn}

\author{Junnan Wu}
\affiliation{%
 \institution{MiLM Plus, Xiaomi Inc.}
 \city{Beijing}
 \country{China}}
\email{wujunnan1@xiaomi.com}

\author{Lichun Fan}
\authornote{Corresponding author}
\affiliation{%
  \institution{MiLM Plus, Xiaomi Inc.}
  \city{Beijing}
  \country{China}}
  \email{fanlichun1@xiaomi.com}

\author{Zhenbo Luo}
\affiliation{%
  \institution{MiLM Plus, Xiaomi Inc.}
  \city{Beijing}
  \country{China}}
\email{luozhenbo@xiaomi.com}

\author{Jian Luan}
\affiliation{%
  \institution{MiLM Plus, Xiaomi Inc.}
  \city{Beijing}
  \country{China}}
\email{luanjian@xiaomi.com}

\author{Jing Lu}
\authornotemark[1]
\affiliation{%
  \institution{Nanjing University}
  \city{Nanjing}
  \country{China}}
  \email{lujing@nju.edu.cn}

\renewcommand{\shortauthors}{Wang et al.}

\begin{abstract}
Language model (LM)-based speech enhancement (SE) can generate natural-sounding speech, but under severe noise it often suffers from unreliable conditioning, leading to perceptually plausible yet linguistically incorrect outputs. To address this issue, we propose L3-SE, a noise-invariant acoustic-semantic distillation framework for reducing linguistic hallucination in LM-based SE. The proposed method learns a noise-invariant conditioning encoder from noisy speech by jointly distilling two complementary clean-speech targets: an acoustic target for reconstruction fidelity and a semantic target for linguistic consistency. The resulting noise-invariant acoustic-semantic representations are used to condition a decoder-only autoregressive language model, which predicts clean acoustic tokens that are decoded into enhanced speech. To support high-quality generation, we further employ a high-fidelity codec built on learnable weighted WavLM layer representations as the discrete acoustic interface. By improving the reliability of conditioning under adverse conditions, the proposed framework substantially reduces hallucination and improves content faithfulness. Experiments show that the proposed method consistently outperforms prior LM-based speech enhancement baselines on linguistic consistency metrics, with especially clear gains under low-SNR and reverberant conditions, while maintaining competitive perceptual quality. Audio samples are available at \url{https://max1wz.github.io/L3-SE-Demo-Page/}. The complete source code will be released after the manuscript is accepted.
\end{abstract}




\maketitle

\section{Introduction}

Speech enhancement (SE) aims to recover clean speech from noisy mixtures, improving perceptual quality and intelligibility for both human listening and downstream applications such as speech communication, automatic speech recognition (ASR) \cite{SE_for_ASR3}, and speaker verification \cite{SE_for_SV2}. Deep-learning-based SE methods generally fall into two families: discriminative (or predictive) \cite{TF-GridNet, BSRNN} and generative approaches. While discriminative models can suppress noise effectively, they often over-suppress speech components under severe noise or reverberation, degrading naturalness \cite{FlowSE,SELM}.
In contrast, generative SE produces enhanced speech via generation formulations, including diffusion models \cite{StoRM,CDiffSE}, flow-matching models \cite{FlowSE,lessismore}, and language model (LM)-based approaches \cite{GenSE, LLaSE-G1, UniSE}, demonstrating strong capabilities in synthesizing speech with superior perceptual quality. 

However, generative SE also introduces a fundamental challenge: perceptual plausibility does not necessarily imply content faithfulness \cite{URGENT2025}. When the conditioning information extracted from distorted speech is unreliable, the generative prior may dominate and produce outputs that sound clean but deviate from the original spoken content, for example through word substitutions, deletions, or other semantic distortions. This failure mode, referred to as \emph{linguistic hallucination} \cite{PASE}, becomes especially pronounced under severe noise and reverberation. Therefore, for generative SE, the central issue is not only whether the model can generate natural speech, but whether its conditioning representations remain robust enough to preserve the original linguistic content.

Among generative approaches, LM-based speech enhancement (SE) is particularly promising because autoregressive language models can capture long-range temporal dependencies and high-level linguistic structure. Existing LM-based methods mainly follow either a single-stage or a two-stage design. In single-stage pipelines, degraded-speech representations are directly used to condition an LM for token prediction \cite{LLaSE-G1, UniSE}. While simple and effective, this design exposes the generator to corrupted conditioning, making it more vulnerable to linguistic hallucination under adverse conditions. In two-stage pipelines, the intermediate representation is first refined and then used to guide token generation in a second stage \cite{GenSE, OmniGSE, Genhancer}. This coarse-to-fine strategy suggests that improving conditioning before generation can be beneficial. However, existing methods typically focus on only one type of representation, either acoustic representations derived from neural audio codecs (NACs) \cite{DAC} or semantic representations extracted by ASR or self-supervised learning (SSL) speech encoders \cite{Whisper-Large-v3, WavLM, HuBERT}. This is limiting for hallucination-resistant LM-based SE, since acoustic representations support reconstruction fidelity, speaker characteristics, and prosody, whereas semantic representations are crucial for maintaining linguistic consistency and preventing content drift.

A related line of work improves downstream robustness by learning noise-invariant speech representations from corrupted inputs \cite{Wav2vec-Switch, Noise-robust_wav2vec2.0, PASE, RobustDistiller}. While these studies show that robust upstream representations can substantially benefit downstream performance, they are still typically centered on a single aspect of the speech signal, most often semantic or contextualized representations for robust recognition and invariance. For LM-based SE, however, reducing linguistic hallucination requires both semantic robustness and acoustic robustness. This motivates our design of a noise-invariant encoder jointly constrained by acoustic and semantic targets, rather than optimizing either aspect alone.

We address this problem with \textbf{L3-SE}, a framework for \textbf{L}owering \textbf{L}inguistic hallucination in \textbf{L}M-based \textbf{S}peech \textbf{E}nhancement via noise-invariant acoustic-semantic distillation. Starting from a single SSL encoder, WavLM \cite{WavLM}, we build two task-specialized teacher models, WavCodec and WavS2T, which share the same WavLM backbone but are optimized for different objectives. WavCodec is trained with a speech codec objective, while WavS2T is trained with a speech-to-transcript objective. Rather than manually selecting a fixed backbone layer, each teacher learns its own task-specialized layer aggregation, yielding complementary representations that emphasize either low-level acoustic detail or high-level linguistic content. We then use the encoder parts of WavCodec and WavS2T as the acoustic and semantic teacher encoders, respectively, freeze them, and train a noisy-speech student, NI-Encoder, to match the corresponding clean-speech teacher targets through representation-level alignment losses. NI-Encoder follows the same overall architectural form as the teachers, namely a shared WavLM backbone followed by acoustic and semantic heads, and produces acoustic-semantic representations from the same noisy input. In this way, low-level acoustic fidelity and high-level linguistic consistency are jointly imposed on a single noise-invariant encoder, providing more reliable conditioning for downstream generation.

Built on the learned noise-invariant acoustic-semantic representations, we further employ an LM to generate clean acoustic tokens for speech reconstruction. The resulting L3-SE framework contains three components: (i) NI-Encoder, which produces noise-invariant acoustic-semantic conditioning representations; (ii) an autoregressive LM that predicts clean acoustic tokens; and (iii) the WavCodec decoder, which converts predicted tokens into waveforms. Here, WavCodec primarily serves as a discrete acoustic interface for LM-based generation and waveform reconstruction, while the main contribution of this work lies in improving the robustness and faithfulness of the conditioning pathway through acoustic-semantic distillation. By conditioning the LM on the distilled representations, L3-SE reduces linguistic hallucination and improves content faithfulness while preserving competitive perceptual quality.

The main contributions of this work are summarized as follows:
\begin{itemize}
\item \textbf{Noise-invariant acoustic-semantic distillation for LM-based speech enhancement.} We propose a joint distillation strategy that learns noise-robust conditioning representations by simultaneously matching acoustic and semantic clean-speech targets, improving both reconstruction fidelity and linguistic consistency under adverse conditions.
\item \textbf{A shared-backbone teacher-student design for reducing linguistic hallucination.} We construct acoustic and semantic teachers on top of the same SSL backbone and use them to supervise a student encoder with acoustic and semantic branches, enabling paired conditioning representations for LM-based speech enhancement.
\item \textbf{An LM-based speech enhancement framework with improved content faithfulness.} Built on NI-Encoder and a discrete acoustic interface, L3-SE predicts clean acoustic tokens for waveform reconstruction and effectively reduces linguistic hallucination while maintaining competitive perceptual quality.
\end{itemize}

\section{Related work}

\subsection{LM-based speech enhancement}
Recent works have explored applying LMs to SE, typically by predicting clean speech tokens in a discrete latent space. SELM \cite{SELM} combines WavLM with k-means clustering to obtain discrete speech tokens and then uses an LM to map noisy tokens to clean ones. LLaSE-G1 \cite{LLaSE-G1} instead conditions a decoder-only LM on continuous WavLM representations and predicts discrete clean-speech tokens for waveform reconstruction, while UniSE \cite{UniSE} follows a similar paradigm for autoregressive acoustic token generation.

More recent methods employ staged or hierarchical generation strategies. GenSE \cite{GenSE} adopts a two-stage framework that first predicts enhanced semantic tokens and then generates clean acoustic tokens. OmniGSE \cite{OmniGSE} follows a continuous-to-discrete collaborative design, where pre-quantized codec-domain features are first enhanced in the continuous domain and then used to condition hierarchical RVQ token generation. Masked generation has also been explored for generalized enhancement settings, as in MaskSR \cite{MaskSR} and AnyEnhance \cite{AnyEnhance}. These studies highlight the promise of LM-based SE, but existing methods mainly focus on either semantic refinement or acoustic restoration, whereas jointly learning conditioning representations that are robust in both acoustic and semantic aspects remains less explored.

\subsection{Noise-invariant speech representation}
Another related direction aims to improve downstream robustness by learning noise-invariant speech representations from corrupted inputs. Prior work has explored this idea through consistency-based robust pre-training and distillation-based robustness objectives. For example, \citet{Wav2vec-Switch} augments wav2vec 2.0 \cite{wav2vec2} with an additional contrastive objective on clean--noisy speech pairs by switching their quantized targets, thereby enforcing prediction consistency between clean and noisy conditions. \citet{Noise-robust_wav2vec2.0} similarly uses paired noisy and clean speech: the noisy branch is optimized with clean quantized targets, together with an additional consistency loss between noisy and clean features, to learn more noise-invariant representations for ASR. 

More closely related to our work are distillation-based approaches that explicitly align noisy-speech representations with clean-speech teacher targets. RobustDistiller \cite{RobustDistiller} improves environmental robustness through a teacher--student framework that distills selected intermediate layers from clean speech, and further combines distillation with data augmentation and an auxiliary denoising objective. PASE \cite{PASE} extends this noisy-to-clean distillation idea to generative speech enhancement by adapting a pre-trained WavLM model with a denoising representation distillation objective, using the backbone's final-layer representation as the teacher target. These studies show that robust upstream representations can substantially benefit downstream tasks. However, prior methods are mostly designed for single-space representation enhancement, and typically focus on a single type of teacher target at a time. In contrast, our work targets LM-based speech enhancement and jointly distills acoustic and semantic targets within the same framework, explicitly learning paired noise-invariant conditioning representations for content-faithful generation.

\begin{figure*}[ht]
  \begin{center}
    \centerline{\includegraphics[width=1.55\columnwidth]{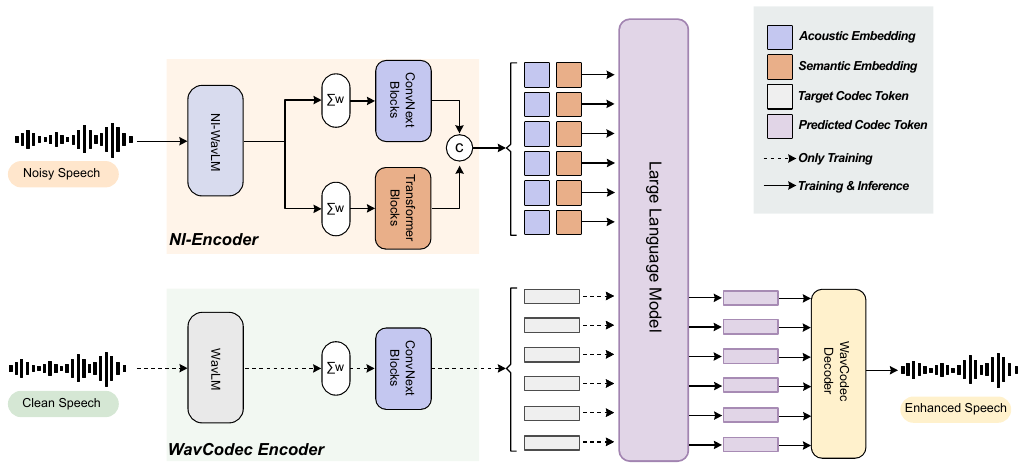}}
    \caption{
      Overview of L3-SE. NI-Encoder extracts acoustic and semantic representations from noisy speech, which are projected as prefix embeddings. LLM predicts clean WavCodec tokens, which are decoded to reconstruct enhanced speech. 
    }
    \label{architecture}
  \end{center}
\end{figure*}

\section{Proposed method}

\subsection{Overall architecture}

In this section, we present L3-SE, an LM-based speech enhancement framework designed to reduce linguistic hallucination by improving the reliability of conditioning representations under noisy conditions. As illustrated in Figure \ref{architecture}, the framework consists of three components: (i) NI-Encoder, which produces paired noise-invariant acoustic and semantic representations from noisy speech; (ii) a decoder-only autoregressive LM that predicts clean acoustic tokens conditioned on these representations; and (iii) the WavCodec decoder, which converts acoustic tokens into waveforms. Among these components, NI-Encoder is the core contribution of our framework. It adopts the same overall architectural form as the teacher encoders, namely a shared WavLM backbone followed by two task-specific heads: an acoustic head implemented with ConvNeXt \cite{ConvNext} blocks and a semantic head implemented with Transformer \cite{Transformer} blocks. Given noisy input speech, NI-Encoder produces paired acoustic and semantic representations that are explicitly regularized through acoustic-semantic joint distillation, yielding more reliable conditioning signals for downstream generation. The resulting acoustic and semantic representations are projected by lightweight adapters and concatenated along the feature dimension to form the conditioning prefixes of the LM. Based on these prefixes, the LM autoregressively predicts clean acoustic tokens, which correspond to the tokens extracted by the WavCodec encoder from clean reference speech during training. Finally, the predicted acoustic tokens are decoded by the WavCodec decoder to reconstruct the enhanced speech. In this way, L3-SE reduces linguistic hallucination by strengthening the conditioning pathway, while leveraging acoustic token generation for high-fidelity waveform reconstruction.

\subsection{Joint Acoustic-semantic representation distillation}

The SSL encoder WavLM provides a hierarchy of hidden representations, where different layers encode complementary aspects of speech \cite{WavLM}. Prior analyses show that lower layers mainly preserve local acoustic structure, whereas upper layers contain richer phonetic and linguistic abstractions \cite{Phonetic_analysis, Comparative_layerwise_SSL_analysis}. A standard way to exploit this hierarchy is to learn a weighted sum over layers. However, for speech enhancement, optimizing a single layer mixture directly for reconstruction often biases the representation toward early layers \cite{DeVo}, which weakens the semantic cues needed to reduce linguistic hallucination.

\begin{figure}[t]
  \begin{center}
    \centerline{\includegraphics[width=0.75\columnwidth]{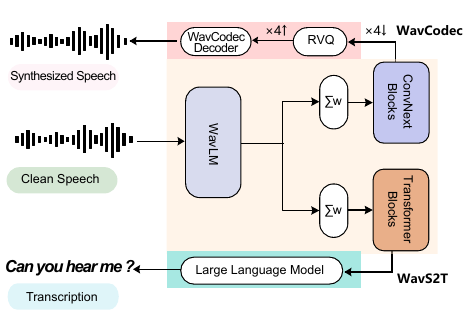}}
    \caption{
      Illustration of WavCodec and WavS2T.
    }
    \label{teacher_fig}
  \end{center}
\end{figure}

To address this issue, as illustrated in Figure \ref{teacher_fig}, we construct two task-specialized teachers on top of a shared pretrained WavLM backbone and train them with complementary objectives: WavCodec for acoustic reconstruction and WavS2T for speech-to-text modeling. Each teacher learns its own layer-mixture weights and task-specific head, producing one representation specialized for acoustic fidelity and another specialized for linguistic consistency. These clean-speech teacher representations are then used as supervision targets for the student NI-Encoder, which is trained on noisy inputs.

For the acoustic teacher, we develop WavCodec, a neural codec based on a VQGAN \cite{VQGAN} formulation. It takes a learnable weighted sum of WavLM layer representations as input and adopts an improved Vocos \cite{Vocos} backbone following WavTokenizer \cite{WavTokenizer}, together with temporal downsampling and upsampling blocks to control the frame rate. The encoder output is downsampled to 12.5~Hz and quantized with RVQ using 8 codebooks of size 1024, resulting in a total token rate of 100. The quantized latents are then decoded to reconstruct the waveform. During training, we train WavCodec with the standard codec objective used in DAC \cite{DAC}, including multi-scale mel reconstruction, codebook and commitment losses, as well as adversarial and feature-matching losses.

For the semantic teacher, we train WavS2T with a speech-to-transcript objective to encourage the model to emphasize linguistic information in WavLM representations. As with WavCodec, we first compute a learnable weighted sum over WavLM layers, and then apply Transformer blocks to obtain continuous semantic representations. These representations are projected into the hidden space of a frozen text LLM (Qwen3-0.6B \cite{Qwen3}) and used as prefix embeddings to predict transcript tokens under a next-token prediction (NTP) loss. This yields semantic representations that are informative of transcript content and compatible with downstream decoder-only LM conditioning \cite{DualSpeechLM, MiMo}.

\begin{figure}[t]
  \begin{center}
    \centerline{\includegraphics[width=0.75\columnwidth]{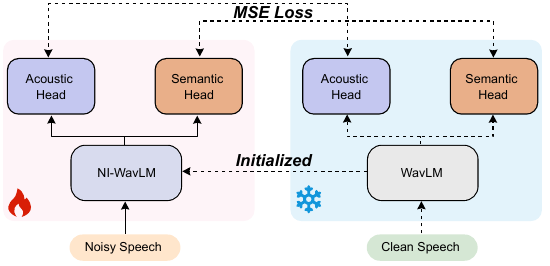}}
    \caption{
      Acoustic-semantic joint distillation for learning NI-Encoder. A frozen teacher (right) extracts acoustic and semantic targets from clean speech, and a student (left) is trained to match both targets from noisy input.
    }
    \label{distill fig}
  \end{center}
\end{figure}

With the two teachers trained, we learn NI-Encoder by acoustic-semantic joint distillation as illustrated in Figure \ref{distill fig}. The acoustic teacher is the encoder of WavCodec, and the semantic teacher is the encoder of WavS2T. The student NI-Encoder follows the same architecture, with a shared WavLM backbone and two task-specific heads, while all teacher parameters remain frozen. Given a paired noisy-clean utterance $(\hat{x}, x)$, the student takes the noisy speech $\hat{x}$ as input and predicts acoustic and semantic representations, while the frozen teachers extract the corresponding targets from the clean speech $x$. 

During training, we optimize the student by minimizing feature-level mean-squared error (MSE) losses for the two branches:

\begin{align}
\label{distill eq}
\mathcal{L}_{\text{distill}} = \mathrm{MSE}\!\left(E^{A}_{S}(\hat{x}),\, E^{A}_{T}(x)\right) + \mathrm{MSE}\!\left(E^{S}_{S}(\hat{x}),\, E^{S}_{T}(x)\right),
\end{align}

where $E^{A}_{T}(\cdot)$ and $E^{S}_{T}(\cdot)$ denote the frozen acoustic and semantic teacher encoders evaluated on the clean speech $x$, while $E^{A}_{S}(\cdot)$ and $E^{S}_{S}(\cdot)$ denote the corresponding acoustic and semantic branches of the student evaluated on the noisy input $\hat{x}$. This joint formulation encourages the learned conditioning space to be both noise-invariant and content-faithful.

\subsection{Autoregressive LM}

Following recent speech generation systems that adapt pretrained decoder-only LLMs for speech token modeling \cite{Cosyvoice2, sparktts, StarVC}, we use Qwen3-0.6B \cite{Qwen3} as the autoregressive backbone and fine-tune it to predict clean acoustic tokens. The LM is conditioned on the noise-invariant acoustic and semantic representations produced by NI-Encoder. Two lightweight adapters project these representations into the LM hidden space, and the projected sequences are concatenated along the feature dimension to form the conditioning prefix $\mathbf{p}$. Conditioned on $\mathbf{p}$, the LM autoregressively predicts the clean WavCodec token sequence, which is then decoded into the enhanced waveform.

Because WavCodec employs RVQ with multiple codebooks, we serialize the target tokens using a flattening scheme \cite{MusicLM} and train the LM with a shared output head. This provides a simple and effective interface for autoregressive token modeling in our setting.

During training, we use teacher forcing \cite{teacherforcing} strategy and NTP loss:
\begin{equation}
\mathcal{L}_{\text{LM}}
= -\frac{1}{TK}\sum_{n=1}^{TK}
\log p_{\phi}\!\left(y_{n}\mid y_{<n}, \mathbf{p}\right),
\end{equation}
where $\phi$ denotes the trainable parameters of the language model, $\mathbf{p}$ denotes the concatenated prefix embeddings, and $\mathbf{y}=(y_1,\ldots,y_{TK})$ is the flattened clean token sequence, with $T$ denoting the number of codec frames and $K$ the number of RVQ codebooks. This objective trains the LM to map noisy-speech conditioning prefixes to clean acoustic token trajectories.

\section{Experiments}

\subsection{Datasets}
\label{datasets}
We use clean speech from the English (EN) and Chinese (ZH) subsets of Emilia \cite{emilia}. For training WavCodec, we use a filtered subset of approximately 4.3k hours with DNSMOS $> 3.50$; for other models, we use a larger clean-speech subset of approximately 22k hours with DNSMOS $> 3.40$. Noise clips are drawn from DNS5 \cite{DNS5}, WHAM! \cite{WHAM}, FSD50K \cite{FSD50K}, and FMA \cite{FMA}, and room impulse responses (RIRs) are sampled from OpenSLR26 and OpenSLR28 \cite{openSLR}. Training mixtures are generated on the fly: for each example, a clean utterance is convolved with a randomly selected RIR with probability 0.5 and then mixed with a randomly sampled noise clip at an SNR uniformly sampled from $[-5, 15]$~dB. The clean utterance is used as the training target. All audio is resampled to 16~kHz.

For evaluation, we use the official DNS1 test set \cite{DNS1}, which contains two subsets, \textit{no-reverb} and \textit{with-reverb}, each with 150 utterances. In addition, we construct a simulated evaluation set with transcripts for controlled WER analysis. Clean speech is drawn from the LibriSpeech \textit{test-clean} split \cite{LibriSpeech}; after duration filtering, we select 913 utterances between 5 and 10 seconds. Noise segments are sampled from the validation portion of URGENT 2025 \cite{URGENT2025} to avoid overlap with training data. Using the same mixing procedure as in training, we generate two evaluation regimes: a \textit{general-SNR} set with SNR uniformly sampled from $[-5, 15]$~dB, and a \textit{low-SNR} set with SNR uniformly sampled from $[-5, 5]$~dB. Compared with DNS1, these simulated sets provide more samples, broader and more controlled SNR coverage, and reference transcripts for WER evaluation. 

\subsection{Baselines}

\textbf{Speech codec baselines.} We compare WavCodec against representative low-bitrate neural codecs and tokenizers. These baselines can be roughly divided into two groups. The first group does not explicitly introduce semantic supervision, including SimCodec \cite{GenSE}, BigCodec \cite{BigCodec}, and WavTokenizer \cite{WavTokenizer}. The second group incorporates semantic constraints from pretrained speech models, including Mimi \cite{Moshi}, XY-Tokenizer \cite{XY-Tokenizer}, X-codec2 \cite{Llasa}, and BiCodec \cite{sparktts}. While high-fidelity codecs such as DAC \cite{DAC} and EnCodec \cite{Encodec} can achieve strong reconstruction quality at higher bitrates, prior work has shown that their performance degrades sharply as bitrate is reduced \cite{sparktts, Llasa}; therefore, we do not include them as low-bitrate baselines in our comparisons. For all codec baselines, we use the official released checkpoints for evaluation.

\textbf{Speech enhancement baselines.} We compare against representative SE systems from three categories. The first category contains discriminative models, including TF-GridNet \cite{TF-GridNet} and BSRNN \cite{BSRNN}. The second category contains generative models, including the vocoder-based model PASE \cite{PASE} and the flow-matching-based models SenSE \cite{SenSE} and BSRNN-Flow \cite{lessismore}. The third category contains LM-based generation models, including LLaSE-G1 \cite{LLaSE-G1}, UniSE \cite{UniSE}, and AnyEnhance \cite{AnyEnhance}. For all SE baselines, we use official released checkpoints for evaluation.


\subsection{Evaluation metrics}
We adopt multiple evaluation metrics to comprehensively assess performance from three perspectives. 




\begin{itemize}
\item \textbf{Perceptual quality metrics}: DNSMOS \cite{DNSMOS-P835} and UTMOS \cite{UTMOS}, which evaluate perceptual speech quality without requiring clean references. 
\item \textbf{Linguistic-consistency metrics}: Levenshtein phoneme similarity (LPS) \cite{LPS}, SpeechBERTScore (SBS) \cite{SpeechBERTScore}, and word error rate (WER), which measure how faithfully the enhanced speech preserves the linguistic content of the corresponding clean reference. In our implementation, SBS is computed using a HuBERT-based \cite{HuBERT} model, LPS is computed using a wav2vec 2.0-based \cite{wav2vec2} model, and WER is computed using Whisper large-v3 \cite{Whisper-Large-v3}. 
\item \textbf{Speaker-preservation metrics}: speaker similarity (SIM), which reflects the preservation of speaker-related characteristics and is computed using a WavLM-based \cite{WavLM} speaker verification model.
\end{itemize}

For the speech codec task, we report PESQ \cite{PESQ}, STOI \cite{ESTOI}, SIM, and WER. For the speech enhancement task, we report DNSMOS, UTMOS, LPS, SBS, SIM, and WER. For DNS1, where reference transcripts are unavailable, we use the transcription of the corresponding clean reference speech as a pseudo-label and report dWER, defined as the WER difference between the enhanced speech and its clean reference. Unless otherwise stated, higher is better for all metrics except WER and dWER.

\begin{table*}[t]
\caption{Main training hyperparameters of different model components.}
\begin{tabular}{lcccccc}
\toprule
 &
Batch Size &
Warmup steps &
Total steps &
Minimum learning rate &
Peak learning rate \\
\midrule

WavCodec-Stage1 & 80 & 10k & 100k & 1e-6 & 1e-4\\
WavCodec-Stage2 & 20 & 5k & 50k & 1e-6 & 2e-5\\
WavS2T & 16 & 20k & 200k & 1e-6 & 1e-4\\
Noise-Invariant Encoder & 6 & 20k & 200k & 1e-6 & 1e-4\\
Language Model & 6 & 5k & 100k & 1e-5 & 1e-4\\
\bottomrule
\end{tabular}
\label{training}
\end{table*}

\subsection{Implementation Details}
\textbf{WavCodec.} We use WavLM-Large as the shared encoder backbone throughout the framework. For the acoustic teacher WavCodec, both the encoder and decoder are built on ConvNeXt blocks. The encoder contains 6 ConvNeXt blocks and the decoder contains 12 ConvNeXt blocks, both operating on 1024-dimensional features with an intermediate dimension of 2048. Two successive $\times 2$ downsampling blocks are applied before RVQ quantization to reduce the frame rate from 50~Hz to 12.5~Hz, and two corresponding $\times 2$ upsampling blocks are used after quantization to restore the temporal resolution. The quantizer uses 8-layer RVQ with 1024 entries per codebook. During adversarial training, we use a multi-period discriminator (MPD) \cite{HiFiGAN} and a multi-band multi-scale STFT discriminator, following DAC \cite{DAC} codec training.

\textbf{WavS2T.} For the semantic teacher WavS2T, we use a 6-layer Transformer encoder with hidden dimension 1024, feed-forward dimension 4096, and 16 attention heads. Its output is projected to a frozen Qwen3-0.6B text LLM to optimize the transcription-oriented objective. The student NI-Encoder follows the same overall architecture as the teachers, using a ConvNeXt-based acoustic head and a Transformer-based semantic head on top of the shared WavLM backbone.

\textbf{LM.} We fine-tune a Qwen3-0.6B LM for downstream acoustic token generation. The acoustic and semantic adapters project 1024-dimensional representations to the LM hidden space, and the output head predicts over the flattened RVQ vocabulary corresponding to the 8-codebook acoustic tokenization.

All models are trained with AdamW \cite{AdamW} on 8 GPUs using a linear warmup followed by cosine decay. WavCodec is trained in two stages: the first stage optimizes the generator only, while the second stage jointly optimizes the generator and discriminator. During WavCodec and WavS2T training, the WavLM backbone is kept frozen. During acoustic-semantic joint distillation, all parameters of NI-Encoder, including the WavLM backbone, are updated. For LM training, NI-Encoder is frozen and the backbone is fine-tuned with teacher forcing. The main training hyperparameters are summarized in Table \ref{training}. Additional implementation details are provided in the supplementary material. The code will be released upon publication.

\begin{table*}[t]
  \caption{Comparison results on LibriSpeech \textit{test-clean}. Semantic Sup. denotes whether semantic supervision is used, and Token Rate denotes the number of tokens per second. \textbf{Bold} and \underline{underline} indicate the best and second-best results, respectively.}
  \label{Codec}
\begin{tabular}{lccccccccc}
\toprule
Model &
Codebook Size &
Semantic Sup. &
Frame Rate &
Token Rate &
BPS &
STOI $\uparrow$ &
PESQ $\uparrow$ &
SIM $\uparrow$&
WER $\downarrow$\\
\midrule

\textbf{Clean} & -- & -- & -- & -- & -- & 1.00 & 4.64 & 1.00 & 2.19\\
\midrule
SimCodec      & 8192 &  No  & 100 & 100 &1300   & \underline{0.93} & \textbf{2.70} & \underline{0.82} & 2.84\\
BigCodec     & 8192 &  No    &  80 & 80 &1040   & \textbf{0.94} & 2.68  & \textbf{0.84} & 2.91\\
WavTokenizer   & 4096   & No  & 75  & 75 &900     & 0.90 & 2.13  & 0.65 & 4.15\\

\midrule
Mimi           & 2048 &  Yes  & 12.5 & 100 & 1100  &    0.91 & 2.26  & 0.74  & 3.24\\
XY-Tokenizer      & 1024 &  Yes   & 12.5 & 100 &1000    &  0.92 & 2.49 & \textbf{0.84} & \textbf{2.46}\\
X-codec2      & 65536  & Yes  & 50 & 50 & 800     & 0.92 & 2.43  & \underline{0.82}  & 2.61\\
BiCodec      & 8192  & Yes & 50 & 50 & 650   &0.92 & 2.51   & 0.80  & 2.72\\
\midrule

\textbf{WavCodec} & 1024 & No & 12.5 & 100 & 1000    & \textbf{0.94} & \underline{2.69} & \textbf{0.84} & \underline{2.52} \\
\bottomrule
\end{tabular}
\end{table*}

\subsection{Speech codec results}


Table \ref{Codec} reports the reconstruction performance of WavCodec and several low-bitrate codec baselines on LibriSpeech \textit{test-clean}. Overall, WavCodec achieves a strong balance between acoustic fidelity and linguistic consistency. It attains the highest STOI and SIM, while also reaching the second-best PESQ and WER, indicating that it preserves intelligibility, speaker characteristics, and linguistic content well.

Compared with codecs without explicit semantic supervision, WavCodec performs consistently better and shows clear gains in both reconstruction quality and linguistic consistency performance. Compared with codecs that incorporate semantic supervision, WavCodec remains highly competitive: although some baselines achieve slightly lower WER, WavCodec provides stronger or comparable results on STOI, PESQ, and SIM. These results suggest that WavCodec offers a favorable trade-off between acoustic fidelity and linguistic consistency, making it a suitable acoustic token interface for downstream LM-based speech enhancement.

\subsection{Ablation study}
\label{ablation_section}
\begin{figure}[ht]
  \begin{center}
    \centerline{\includegraphics[width=0.8\columnwidth]{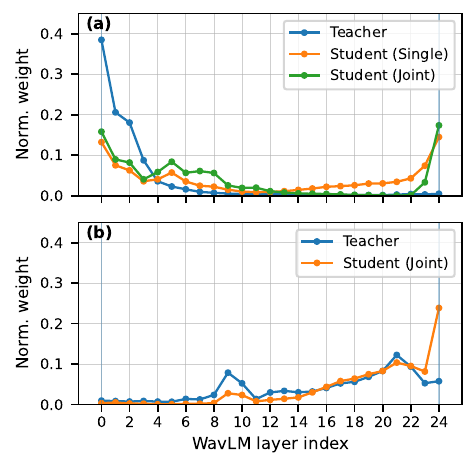}}
    \caption{
      (a) Acoustic layer weights for the teacher, single-distilled student, and joint-distilled student.
  (b) Semantic layer weights for the teacher and joint-distilled student.
    }
    \label{wavlm_weights}
  \end{center}
\end{figure}

\begin{table*}[t]
\caption{Ablation of acoustic-semantic distillation. “Teacher” denotes the clean-speech supervision used for NI-Encoder: A for acoustic teacher, A+S for joint acoustic-semantic teachers, L0+L24 for fixed WavLM layer targets, and A+S-CTC for a CTC-trained semantic teacher. “Prefix” denotes the conditioning prefix fed to the LM. Bold indicates the best result.}
  \label{ablation}
\begin{tabular}{lccccccccc}
\toprule
Exp. & Distillation & Teacher & Prefix & DNSMOS $\uparrow$& UTMOS $\uparrow$& SBS $\uparrow$& LPS $\uparrow$& SIM $\uparrow$& WER $\downarrow$\\
\midrule
(a)  & None & -- & A & 3.19 & 3.04 & 0.772 & 0.730 & 0.917 & 26.2  \\
\midrule
(b) & Single & A & A & 3.18 & 3.28 & 0.839 & 0.856 & 0.946 & 14.7   \\
(c)  & Joint & A+S & A & 3.18 & 3.32 & 0.864 & 0.904 & 0.952 & 8.65 \\
(d)  & Joint & L0+L24 & L0+L24 & 3.17 & \textbf{3.33} & 0.861 & 0.904 & 0.947 & 9.08  \\
(e)  & Joint & A+S-CTC & A+S-CTC & 3.18 & 3.32 & 0.864 & 0.899 & 0.951 & 9.88   \\
(f)  & Joint & A+S & A+S & \textbf{3.19} & \textbf{3.33} & \textbf{0.866} & \textbf{0.909} & \textbf{0.953} & \textbf{7.71}   \\
\bottomrule
\end{tabular}
\end{table*}

Figure \ref{wavlm_weights} and Table \ref{ablation} jointly show how acoustic-semantic distillation reshapes the learned WavLM layer mixtures and how these changes affect enhancement performance. Overall, the results support five observations: (i) encoder distillation is essential, (ii) joint acoustic-semantic distillation is stronger than acoustic-only distillation, (iii) its benefit cannot be explained by selecting only a few salient WavLM layers, (iv) the semantic-teacher objective matters, and (v) using both acoustic and semantic prefixes further improves downstream generation.

\textbf{Effect of encoder distillation.}
Comparing the non-distilled baseline (Exp.~(a)) with its distilled counterpart (Exp.~(b)) shows that representation distillation is a key ingredient for improving robustness under noise. Moving from no distillation to single acoustic distillation yields large gains in all linguistic-consistency metrics, including clear improvements in SBS, LPS, and SIM together with a substantial reduction in WER, while the changes in perceptual quality metrics are relatively modest. This indicates that the primary role of distillation is not merely to improve perceptual quality, but to provide more reliable conditioning representations for preserving linguistic content and speaker-related information.

\textbf{Effect of joint distillation.}
Under the same acoustic-prefix setting, joint distillation (Exp.~(c)) consistently outperforms single acoustic distillation (Exp.~(b)), with clear gains in linguistic-consistency metrics and a large reduction in WER. This improvement is consistent with the layer-weight analysis in Figure \ref{wavlm_weights}. The acoustic teacher is strongly concentrated on the lowest layers, reflecting its reliance on local reconstruction cues, whereas the semantic teacher places most of its mass on upper layers. After joint distillation, the student heads no longer simply mimic these teacher profiles. In particular, the acoustic student shifts more weight toward low and mid-low layers while also increasing the contribution of the top layer, and the semantic student further emphasizes the highest layer. This suggests that joint supervision encourages the student to combine local acoustic evidence with higher-level linguistic cues, resulting in conditioning features that are more robust to noise and more effective for preserving linguistic consistency than acoustic-only distillation. More broadly, the fact that joint supervision reshapes both student heads suggests that the two clean-speech teacher targets are complementary rather than fully disjoint, and that their partially overlapping information can be further integrated in the student under noisy conditions.

\textbf{Effect of constrained layer targets.}
Motivated by the layer-weight plots, we further construct a constrained-target baseline that replaces the learned acoustic-semantic teachers with fixed layer targets, using L0 as the acoustic target and L24 as the semantic target. As shown in Exp.~(d), this restricted setting performs much better than the non-distilled baseline, confirming that salient low- and high-level layers already provide useful supervision. However, it remains inferior to joint distillation with the full learned acoustic-semantic teachers (Exp.~(c) and (f)), especially on the linguistic-consistency metrics. This comparison indicates that the gain of joint distillation cannot be reduced to simply emphasizing two prominent layers. Instead, the learned task-specialized teachers provide richer supervision than fixed layer selection, allowing NI-Encoder to exploit complementary information across the WavLM hierarchy.

\textbf{Effect of semantic-teacher objective.}
We next compare two training objectives for the semantic teacher while keeping the rest of the framework unchanged. As a baseline, we replace the semantic teacher's transcript decoder with a Connectionist Temporal Classification (CTC) head and train it with a CTC objective \cite{CTC}, which is widely used in ASR for frame-to-label alignment. We then apply the same acoustic-semantic joint distillation and use the resulting acoustic+semantic prefixes to condition the LM. Compared with this CTC-based variant (Exp.~(e)), the NTP-based semantic teacher (Exp.~(f)) consistently improves the linguistic-consistency metrics, yielding higher SBS, LPS, and SIM together with a lower WER, while keeping DNSMOS and UTMOS nearly unchanged. This suggests that the advantage of the NTP objective lies mainly in providing more effective semantic supervision for downstream generation rather than in improving perceptual quality. A likely reason is that NTP learns semantic representations that are better aligned with autoregressive modeling, whereas CTC mainly emphasizes framewise transcription alignment and thus offers weaker guidance for LM-conditioned generation.

\textbf{Effect of acoustic-semantic prefixes.}
Finally, we compare whether both student branches should be exposed to the LM after joint distillation. Comparing Exp.~(c) and (f), where the teacher targets are identical and only the LM prefix differs, shows that using both acoustic and semantic prefixes consistently improves the results over using the acoustic prefix alone. Although the gains on the perceptual quality metrics are small, the best overall SBS, LPS, SIM, and WER are obtained only when both prefixes are used. This indicates that the benefit of acoustic-semantic modeling is not exhausted at the encoder level: once joint distillation has made the two branches noise-consistent, explicitly conditioning the LM on both branches still provides complementary information for generation.

\subsection{Speech enhancement results}

\begin{table*}[t]
  \caption{Comparison results on the DNS1 testset. \textbf{Bold} and \underline{underline} indicate the best and second-best results, respectively.}
  \label{DNS1}
\begin{tabular}{lcccccccccccc}
\toprule
& \multicolumn{6}{c}{DNS1 \textit{no-reverb}} & \multicolumn{6}{c}{DNS1 \textit{with-reverb}} \\
\cmidrule(lr){2-7}\cmidrule(lr){8-13}

& DNSMOS $\uparrow$& UTMOS $\uparrow$& SBS $\uparrow$& LPS $\uparrow$& SIM $\uparrow$& dWER $\downarrow$
& DNSMOS $\uparrow$ & UTMOS $\uparrow$ & SBS $\uparrow$ & LPS $\uparrow$ & SIM $\uparrow$ & dWER $\downarrow$\\
\midrule

Noisy  & 2.48 & 2.37 & 0.797 & 0.898 & 0.980 & 3.55 & 1.39 & 1.30 & 0.608 & 0.633 & 0.924 & 10.1 \\
Clean  & 3.28 & 4.14 & 1.00 & 1.00 & 1.00 & 0.00 & 3.28 & 4.14 & 1.00 & 1.00 & 1.00 & 0.00 \\
\midrule
TF-GridNet & 3.34 & 3.86 & 0.909 & 0.965 & \textbf{0.991} & 3.18 & 2.63 & 1.42 & 0.770 & 0.877 & 0.955 & 8.45 \\
BSRNN & 3.28 & 3.85 & \textbf{0.925} & \textbf{0.972} & \textbf{0.991} & \underline{2.93} & 2.60 & 1.46 & 0.793 & 0.890 & \underline{0.958} & \textbf{7.65} \\
\midrule
PASE & 3.35 & 3.61 & \underline{0.922} & \underline{0.968} & 0.987 & \textbf{2.78} & 2.82 & 1.61 & 0.808 & 0.896 & 0.957 & 9.59 \\
SenSE & 3.38 & 3.85 & 0.916 & 0.967 & \underline{0.989} & 5.49 & 3.37 & 3.55 & \textbf{0.850} & \underline{0.916} & \textbf{0.970} & 11.3 \\
BSRNN-Flow & 3.34 & 3.92 & 0.900 & 0.960 & \textbf{0.991} & 4.67 & 2.54 & 1.43 & 0.750 & 0.848 & 0.953 & 12.9 \\
\midrule
AnyEnhance & \underline{3.42} & \underline{3.95} & 0.906 & 0.959 & 0.988 & 4.52 & 3.20 & 2.75 & 0.798 & 0.871 & 0.955 & 14.0 \\
LLaSE-G1   & 3.14 & 3.23 & 0.810 & 0.893 & 0.953 & 8.77 & 3.05 & 2.35 & 0.681 & 0.676 & 0.902 & 38.9 \\
UniSE     & \underline{3.42} & \textbf{4.06} & 0.877 & 0.924 & 0.977 & 7.78 & \underline{3.41} & \textbf{3.78} & 0.748 & 0.742 & 0.947 & 34.2 \\
\midrule
\textbf{L3-SE} & \textbf{3.44} & 3.86 & 0.910 & 0.961 & 0.985 & 3.45 & \textbf{3.43} & \underline{3.59} & \underline{0.846} & \textbf{0.917} & 0.956 & \underline{8.42} \\
\bottomrule
\end{tabular}
\end{table*}

\begin{table*}[t]
\caption{Comparison results on the simulated testset. \textbf{Bold} and \underline{underline} indicate the best and second-best results, respectively.}
  \label{Libri}
\begin{tabular}{lcccccccccccc}
\toprule
& \multicolumn{6}{c}{Librispeech \textit{general-SNR}} & \multicolumn{6}{c}{Librispeech \textit{low-SNR}} \\
\cmidrule(lr){2-7}\cmidrule(lr){8-13}

& DNSMOS $\uparrow$& UTMOS $\uparrow$& SBS $\uparrow$& LPS $\uparrow$& SIM $\uparrow$& WER $\downarrow$
& DNSMOS $\uparrow$& UTMOS $\uparrow$& SBS $\uparrow$& LPS $\uparrow$& SIM $\uparrow$& WER $\downarrow$\\
\midrule

Noisy  & 1.68 & 1.65 & 0.672 & 0.729 & 0.940 & 7.68 & 1.47 & 1.47 & 0.568 & 0.563 & 0.905 & 12.0 \\
Clean  & 3.08 & 3.47 & 1.00 & 1.00 & 1.00 & 2.07 & 3.08 & 3.47 & 1.00 & 1.00 & 1.00 & 2.07 \\
\midrule
TF-GridNet & 3.16 & 3.07 & 0.869 & 0.925 & \underline{0.977} &7.40 & 3.11 & 2.83 & 0.814 & 0.877 & 0.960 & 11.3  \\
BSRNN & 3.20 & 3.35 & 0.889 & 0.940 & \textbf{0.981} & \underline{7.07} & 3.17 & 3.18 & 0.842 & 0.904 & \textbf{0.969} & 10.0 \\
\midrule
PASE & 3.21 & 3.77 & \textbf{0.898} & 0.942 & 0.975 & 7.68 & 3.17 & 3.05 & \textbf{0.874} & 0.915 & \underline{0.967} & \underline{9.91} \\
SenSE & \textbf{3.42} & \underline{3.91} & 0.871 & \underline{0.945} & 0.973 & 7.16 & \textbf{3.42} & \underline{3.93} & 0.846 & \underline{0.919} & 0.964 & 10.0 \\
BSRNN-Flow & 3.20 & 3.21 & 0.845 & 0.897 & 0.975 & 11.0 & 3.16 & 3.02 & 0.785 & 0.832 & 0.959 & 16.7 \\
\midrule
AnyEnhance & 3.37 & 3.80 & 0.853 & 0.915 & 0.972 & 9.55 & 3.32 & 3.68 & 0.807 & 0.855 & 0.954 & 15.9 \\
LLaSE-G1   & 2.98 & 2.91 & 0.745 & 0.765 & 0.927 & 22.9 &2.96 & 2.75 & 0.693 & 0.672 & 0.906 & 34.9  \\
UniSE     & \underline{3.39} & \textbf{4.06} & 0.811 & 0.829 & 0.949 & 21.1 & \underline{3.40} & \textbf{4.01} & 0.756 & 0.731 & 0.932 & 33.6  \\
\midrule

\textbf{L3-SE} & 3.38 & 3.71 & \underline{0.891} & \textbf{0.949} & 0.972 & \textbf{4.96} & 3.38 & 3.67 & \underline{0.862} & \textbf{0.929} & 0.963 & \textbf{7.13} \\
\bottomrule
\end{tabular}
\end{table*}

Table \ref{DNS1} reports results on the DNS1 test set under \textit{no-reverb} and \textit{with-reverb} conditions. A clear pattern is that discriminative baselines are relatively stable in linguistic-consistency metrics, but their perceptual quality is less favorable in the reverberant condition. On \textit{with-reverb}, TF-GridNet and BSRNN achieve comparatively low dWER, but their DNSMOS and UTMOS remain substantially below most generative systems. In contrast, generative models generally obtain better perceptual quality, but their linguistic consistency is much less stable. Notably, among all generative methods, only PASE and L3-SE achieve lower dWER than the noisy input on both DNS1 subsets, showing that most other generative systems still amplify linguistic deviations rather than correcting them.

This difference becomes more pronounced on the \textit{with-reverb} subset, where reverberation makes conditioning much less reliable. In this more challenging setting, L3-SE achieves the best DNSMOS, the best LPS, and the lowest dWER among the generative methods. Its dWER is second only to BSRNN, while its perceptual quality is far stronger than BSRNN in both DNSMOS and UTMOS. This result is important for our setting: it shows that L3-SE narrows the linguistic-consistency gap, while retaining the perceptual advantage of generative modeling. More broadly, the DNS1 results indicate that reverberation is a particularly difficult regime for LM-based generation, and that improving the noise robustness of acoustic-semantic conditioning is critical for reducing hallucination in such cases.

Table \ref{Libri} further evaluates the models on our simulated LibriSpeech test sets with uniformly sampled SNR in two regimes. This controlled setup provides broader coverage and a more uniform SNR distribution than DNS1. Across both \textit{general-SNR} and \textit{low-SNR}, L3-SE achieves the lowest WER among all compared methods, while also maintaining strong SBS, LPS, and SIM. In the \textit{general-SNR} regime, L3-SE already provides the best WER and the best LPS, together with competitive perceptual quality. In the more challenging \textit{low-SNR} regime, its advantage becomes even clearer: L3-SE again achieves the lowest WER and the strongest overall linguistic-consistency metrics, showing that the proposed conditioning strategy remains effective when the input is heavily corrupted.

The LM-based baselines further highlight this trend. Except for AnyEnhance, which benefits from semantic alignment, LLaSE-G1 and UniSE show severe degradation in linguistic consistency, with especially large WER under \textit{with-reverb} and \textit{low-SNR}. Even when some of these systems obtain favorable perceptual scores, their linguistic-consistency metrics deteriorate sharply, indicating pronounced linguistic hallucination once the conditioning signal becomes unreliable. By contrast, L3-SE remains comparatively stable across all evaluation settings, and its advantage is most evident precisely in the more difficult reverberant and low-SNR conditions. Overall, these results show that L3-SE provides a better balance between perceptual quality and linguistic consistency than competing generative and LM-based baselines, and that noise-invariant acoustic-semantic conditioning is particularly effective for improving robustness in complex scenarios.

\section{Conclusion}
In this work, we presented \textbf{L3-SE}, an LM-based speech enhancement framework designed to reduce linguistic hallucination by improving conditioning reliability under noisy conditions. Our method is built on a noise-invariant acoustic-semantic joint distillation strategy, which learns paired acoustic and semantic representations from noisy speech. Conditioned on these representations, an LM predicts clean acoustic tokens for speech reconstruction. Experiments show that L3-SE achieves stronger linguistic consistency than competing baselines while maintaining competitive perceptual quality, especially under reverberant and low-SNR conditions.

\begin{acks}
All datasets used in this work are publicly available and were used strictly for academic research purposes in accordance with their respective licenses and terms of use. We strictly adhere to the terms, conditions, and usage policies specified by the respective dataset providers. 

We also express our gratitude to the creators and maintainers of these datasets for their valuable contributions to the research community. Their efforts have made this study possible.
\end{acks}


\newpage
\bibliographystyle{ACM-Reference-Format}
\bibliography{ref}

\newpage
\appendix
\onecolumn
\section{Training objective}

\subsection{WavCodec}
\label{training-objective-wavcodec}
For WavCodec, we follow the DAC training objective \cite{DAC}. The generator loss is
\begin{equation}
\mathcal{L}_{G}
= \lambda_{\mathrm{rec}}\,\mathcal{L}_{\mathrm{rec}}
+ \lambda_{\mathrm{adv}}\,\mathcal{L}_{\mathrm{adv}}
+ \lambda_{\mathrm{fm}}\,\mathcal{L}_{\mathrm{fm}}
+ \lambda_{\mathrm{cb}}\,\mathcal{L}_{\mathrm{cb}}
+ \lambda_{\mathrm{com}}\,\mathcal{L}_{\mathrm{com}},
\end{equation}
where we set $\lambda_{\mathrm{rec}}=15$, $\lambda_{\mathrm{fm}}=2$ and $\lambda_{\mathrm{com}}=0.25$, and set $\lambda_{\mathrm{adv}}=\lambda_{\mathrm{cb}}=1$.

The reconstruction loss is an $\ell_1$ multi-scale mel-spectrogram loss:
\begin{equation}
\mathcal{L}_{\mathrm{rec}}
= \sum_{s\in\mathcal{S}} \left\lVert M_s(x) - M_s(\hat{x}) \right\rVert_1,
\end{equation}
where $x$ and $\hat{x}$ denote the target and reconstructed waveforms, respectively.

For adversarial training, we use a multi-period discriminator and a multi-band multi-scale STFT discriminator and denote all discriminators by $\{D_i\}$. We adopt least-squares GAN losses:
\begin{equation}
\mathcal{L}_{D}
= \sum_i \mathbb{E}\Big[(D_i(x)-1)^2 + D_i(\hat{x})^2\Big],
\qquad
\mathcal{L}_{\mathrm{adv}}
= \sum_i \mathbb{E}\Big[(D_i(\hat{x})-1)^2\Big].
\end{equation}
We include a feature matching loss over intermediate discriminator activations:
\begin{equation}
\mathcal{L}_{\mathrm{fm}}
= \sum_i \sum_l \left\lVert D_i^{(l)}(x) - D_i^{(l)}(\hat{x}) \right\rVert_1,
\end{equation}
where $D_i^{(l)}(\cdot)$ denotes the activation at layer $l$ of discriminator $D_i$.

We optimize the RVQ quantizer with codebook and commitment losses:
\begin{equation}
\mathcal{L}_{\mathrm{cb}}
= \left\lVert \mathrm{sg}[\mathbf{z}] - \hat{\mathbf{z}} \right\rVert_2^2,
\qquad
\mathcal{L}_{\mathrm{com}}
= \left\lVert \mathbf{z} - \mathrm{sg}[\hat{\mathbf{z}}] \right\rVert_2^2,
\end{equation}
where $\mathrm{sg}[\cdot]$ denotes the stop-gradient operator. We update discriminator parameters by minimizing $\mathcal{L}_{D}$ and update the generator parameters (encoder, decoder, and quantizer) by minimizing $\mathcal{L}_{G}$.

\subsection{WavS2T}
\label{training-objective-wavs2t}
Given an input utterance $x$ with reference transcript tokens $\mathbf{y}=(y_1,\ldots,y_N)$, the semantic teacher extracts a sequence of continuous speech representations $\mathbf{h}$, which are projected to the text LLM hidden dimension and used as prefix embeddings:
\begin{equation}
\mathbf{p} = P_{\psi}(\mathbf{h}).
\end{equation}
We condition a frozen text LLM Qwen3-0.6B \cite{Qwen3} on $\mathbf{p}$ and train the semantic teacher with teacher forcing by minimizing a next-token prediction (NTP) loss:
\begin{equation}
\mathcal{L}_{\mathrm{NTP}}
= -\frac{1}{N}\sum_{t=1}^{N}\log p_{\phi}\!\left(y_t \mid y_{<t}, \mathbf{p}\right),
\end{equation}
where $p_{\phi}(\cdot)$ denotes the LLM distribution over the next transcript token given the prefix embeddings $\mathbf{p}$ and the preceding tokens $y_{<t}$. During training, we freeze the LLM backbone parameters $\phi$ and optimize only the semantic encoder parameters $\theta$ and projection parameters $\psi$, encouraging $\mathbf{h}$ to be predictive of transcript tokens while preserving the pretrained linguistic prior of the LLM.

\section{Baselines}
\subsection{Speech codec baselines}
We compare WavCodec against representative low-bitrate speech codecs.

\textbf{SimCodec} \cite{GenSE} A single VQ-based codec with codebook reorganization. We use the officially released checkpoint \footnote{\url{https://huggingface.co/yaoxunji/gen-se}} for evaluation.

\textbf{Mimi} \cite{Moshi} An RVQ-based codec with semantic constraint by WavLM . We use the officially released checkpoint \footnote{\url{https://huggingface.co/kyutai/mimi}} for evaluation.

\textbf{XY-Tokenizer} \cite{XY-Tokenizer} An RVQ-based codec with semantic constraint by Whisper \cite{Whisper-Large-v3}. We use the officially released checkpoint \footnote{\url{https://huggingface.co/fdugyt/XY_Tokenizer}} for evaluation.

\textbf{BigCodec} \cite{BigCodec} A single VQ-based codec. We use the officially released checkpoint \footnote{\url{https://huggingface.co/Alethia/BigCodec}} for evaluation.

\textbf{WavTokenizer} \cite{WavTokenizer} A single VQ-based tokenizer. We use the officially released checkpoint \footnote{\url{https://huggingface.co/novateur/WavTokenizer}} for evaluation.

\textbf{X-codec2} \cite{Llasa} A FSQ-based version of X-codec. We use the officially released checkpoint \footnote{\url{https://huggingface.co/HKUSTAudio/xcodec2}} for evaluation.

\textbf{BiCodec} \cite{sparktts} A single VQ based codec with semantic constraint by wav2vec 2.0 \cite{wav2vec2}. We use the officially released checkpoint \footnote{\url{https://huggingface.co/SparkAudio/Spark-TTS-0.5B}} for evaluation.

\subsection{Speech enhancement baselines}
We evaluate against a diverse set of state-of-the-art systems spanning different modeling paradigms.

\textbf{TF-GridNet} \cite{TF-GridNet} A discriminative model that integrates full- and sub-band modeling in the time-frequency domain. We use the released checkpoint provided as the baseline by URGENT 2025 Challenge\footnote{\url{https://huggingface.co/kohei0209/tfgridnet_urgent25}} for evaluation.

\textbf{BSRNN} \cite{BSRNN} A discriminative model that integrates explicit band-splitting in the time-frequency domain. We use the released checkpoint provided as the baseline by URGENT 2026 Challenge\footnote{\url{https://huggingface.co/lichenda/icassp_2026_urgent_baseline}} for evaluation.

\textbf{PASE} \cite{PASE} A generative model that leverages the robust phonological prior embedded in the pre-trained WavLM model to mitigate hallucinations. We use the official inference results provided by the authors for evaluation.

\textbf{AnyEnhance} \cite{AnyEnhance} A generative model based on a masked generative model architecture. We use the official inference results provided by the authors for evaluation.

\textbf{SenSE} \cite{SenSE} A flow-matching-based model that leverages a language model to capture the semantic condition. We use the officially released checkpoint\footnote{\url{https://huggingface.co/ASLP-lab/SenSE}} for evaluation.

\textbf{BSRNN-Flow} \cite{lessismore} A flow-matching-based model that uses BSRNN as the backbone. We use the released checkpoint provided as the baseline by URGENT 2026 Challenge\footnote{\url{https://huggingface.co/lichenda/icassp_2026_urgent_baseline}} for evaluation.

\textbf{LLaSE-G1} \cite{LLaSE-G1} An LM-based model that employs a non-autoregressive (NAR) feature mapping approach. We use the officially released checkpoint\footnote{\url{https://huggingface.co/ASLP-lab/LLaSE-G1}} for evaluation.

\textbf{UniSE} \cite{UniSE} An LM-based model that employs an autoregressive (AR) modeling approach. We use the officially released checkpoint\footnote{\url{https://huggingface.co/QuarkAudio/QuarkAudio-UniSE}} for evaluation.

\section{Testsets}
\begin{figure}[ht]
  \begin{subfigure}[t]{0.3\columnwidth}
    \centering
    \includegraphics[width=\linewidth]{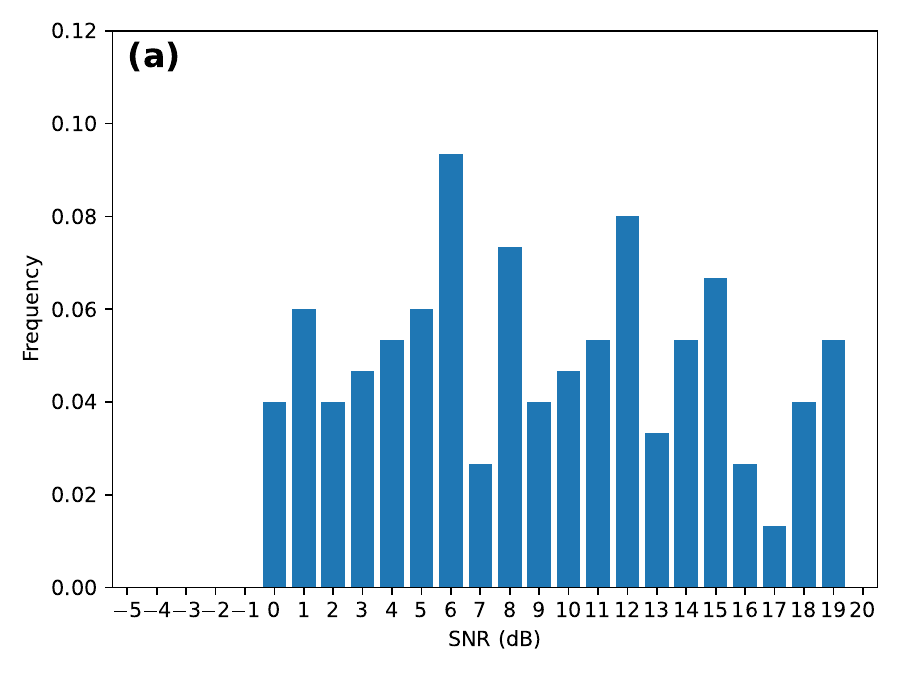}
    \label{dns_hist}
  \end{subfigure}\hfill
  \begin{subfigure}[t]{0.3\columnwidth}
    \centering
    \includegraphics[width=\linewidth]{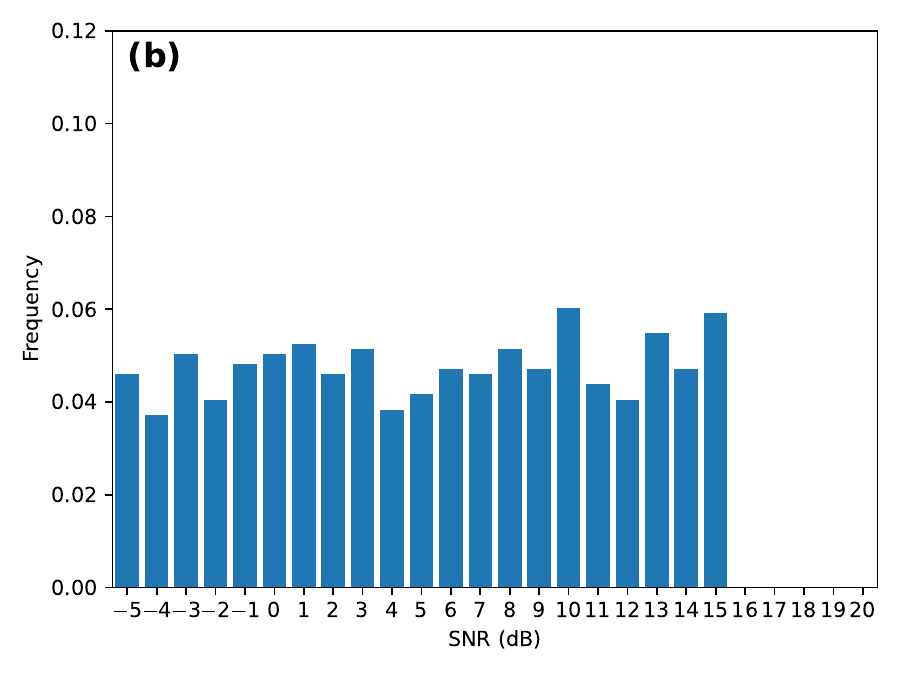}
    \label{libri_normal_hist}
  \end{subfigure}\hfill
  \begin{subfigure}[t]{0.3\columnwidth}
    \centering
    \includegraphics[width=\linewidth]{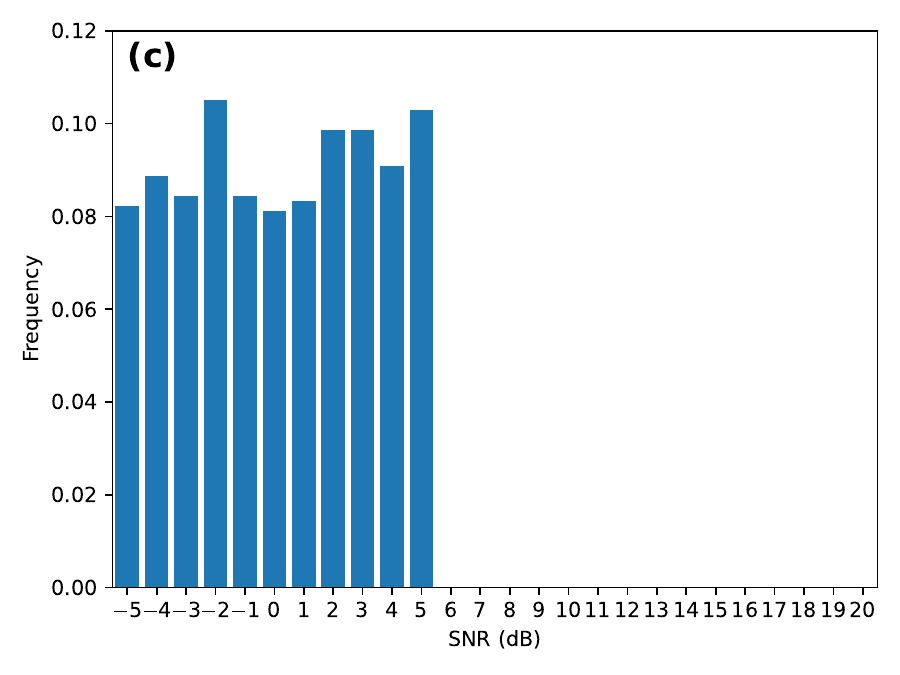}
    \label{libri_low_hist}
  \end{subfigure}
  \caption{Different SNR distributions of DNS1 and our simulated sets. (a) DNS1, (b) general-SNR testset, (c) low-SNR testset}
\label{dataset}
\end{figure}

Relative to the public DNS1 \cite{DNS1} testset, our simulated sets provide substantially more evaluation samples (913 vs. 150), a more controlled and broader SNR coverage, and reference transcripts for WER evaluation. We also visualize and compare the SNR distributions of DNS1 and our simulated sets in Figure \ref{dataset}, showing that DNS1 is skewed toward higher-SNR conditions and exhibits a less uniform distribution.

\section{Audio examples visualization}

\begin{figure*}[ht]
\centerline{\includegraphics[width=0.9\columnwidth]{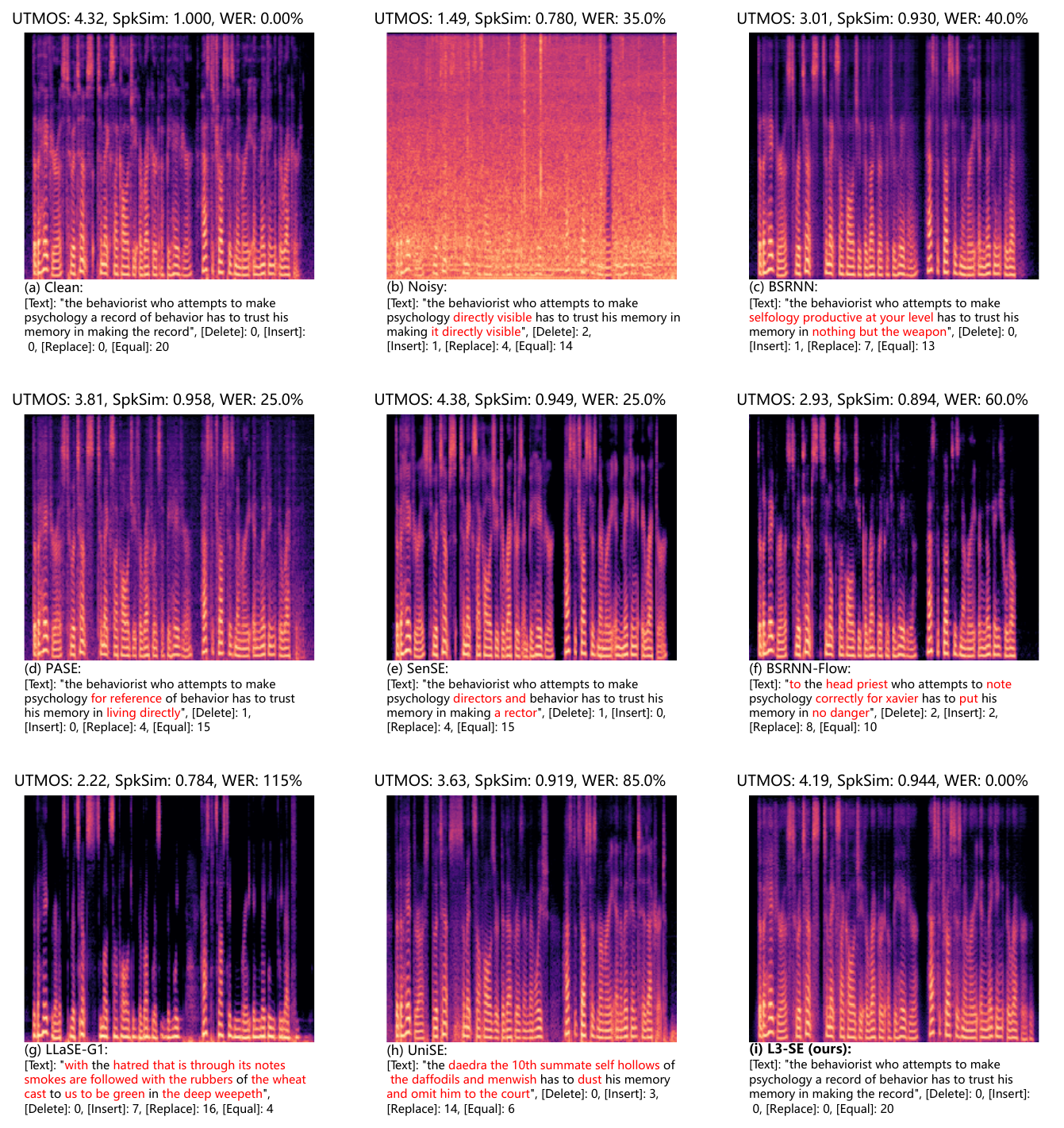}}
    \caption{Spectrogram-based qualitative comparison on a low-SNR utterance. Enhanced spectrograms and ASR transcripts, along with UTMOS, SpkSim, and WER are reported. LM-based baselines (g,h) exhibit severe linguistic hallucinations under low SNR. Our method (i) preserves perceptual quality while maintaining linguistic consistency.
    }
    \label{samples}
\end{figure*}

Figure \ref{samples} provides a qualitative comparison on a representative low-SNR utterance, where the noisy observation is severely corrupted. We report UTMOS and SpkSim as proxies for perceptual quality and speaker preservation, and WER as a measure of linguistic consistency.

In the last row, the LM-based methods in (g) and (h) exhibit pronounced degradation: despite producing plausible-sounding outputs, they introduce substantial lexical substitutions, insertions, and deletions, leading to sharply increased WER and visibly distorted spectro-temporal patterns. Other enhancement baselines ((c)–(f)) display varying degrees of degradation in linguistic fidelity. Some preserve parts of the content but introduce artifacts or partial over-suppression, which is reflected in either reduced UTMOS/SpkSim or increased WER.

In contrast, (i) L3-SE maintains both quality and linguistic consistency. It retains a spectro-temporal structure close to the clean reference while achieving high UTMOS and strong speaker similarity. More importantly, it yields substantially lower WER, indicating improved content preservation. This supports our main claim that learning noise-invariant acoustic–semantic conditioning stabilizes generation under degradation, enabling the model to enhance speech without linguistic hallucinations. 

More examples are available at \url{https://max1wz.github.io/L3-SE-Demo-Page/}. The complete source code will be released after the manuscript is accepted.


\end{document}